\documentclass{vgtc}                          

\ifpdf
  \pdfoutput=1\relax                   
  \usepackage{graphicx}                
  \DeclareGraphicsExtensions{.pdf,.png,.jpg,.jpeg} 
\else
  \usepackage{graphicx}                
  \DeclareGraphicsExtensions{.eps}     
\fi%

\graphicspath{{figures/}{pictures/}{images/}{./}} 

\usepackage{microtype}                 
\PassOptionsToPackage{warn}{textcomp}  
\usepackage{textcomp}                  
\usepackage{mathptmx}                  
\usepackage{times}                     
\usepackage{cite}                      
\usepackage{tabu}                      
\usepackage{booktabs}                  

\newcommand\bparagraph[1]{\vspace{1.0mm}\noindent\textbf{#1}}

\onlineid{0}

\vgtccategory{Research}

\vgtcinsertpkg

\title{Deceiving Audio Design in Augmented Environments : A Systematic Review of Audio Effects in Augmented Reality}

 \author{Esm\'ee Henrieke Anne de Haas\thanks{e-mail: ehadehaas@kaist.ac.kr} %
 \and Lik-Hang Lee\thanks{e-mail:
 likhang.lee@kaist.ac.kr}} %
 \affiliation{\scriptsize Korea Advanced Institute of Science and Technology, Republic of Korea}

 \teaser{
   \centering
   \includegraphics[width=\linewidth]{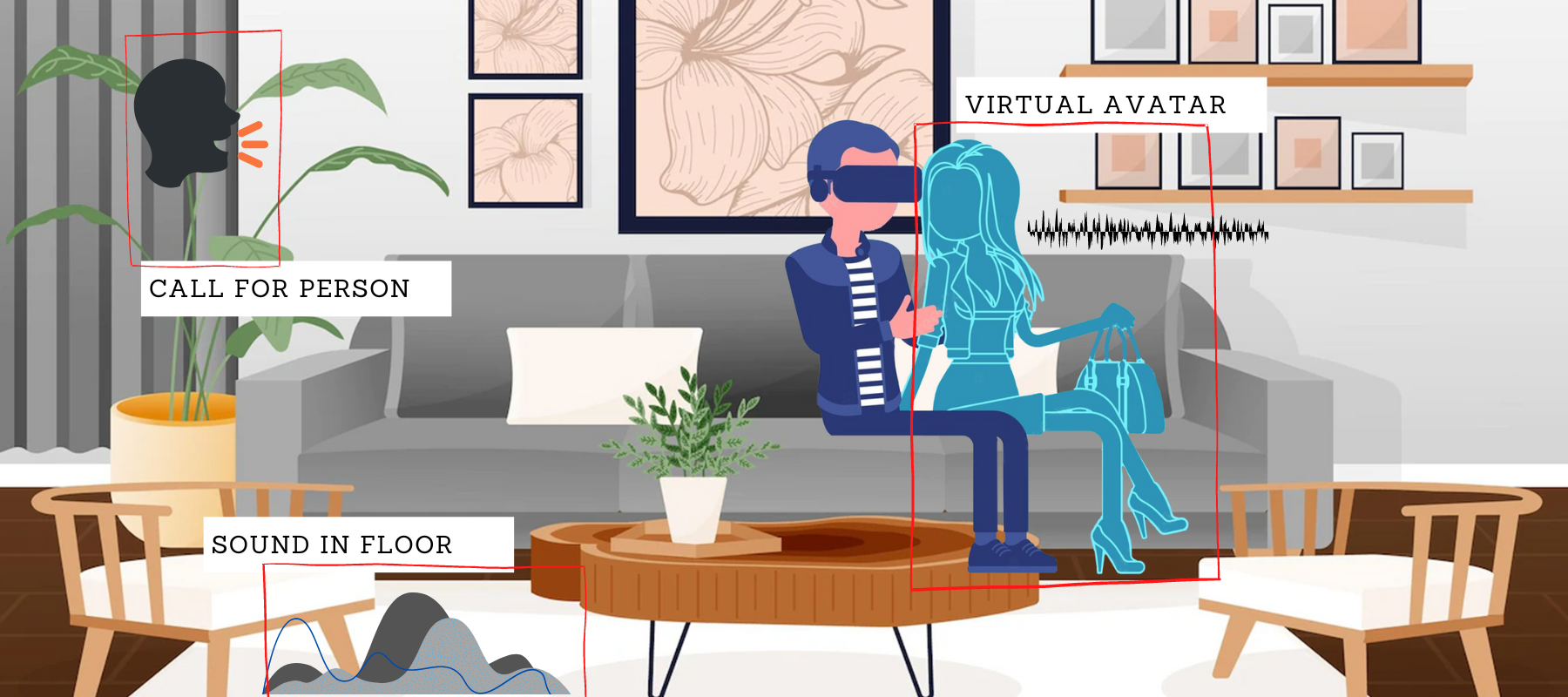}
   \caption{
   Examples of audio distractions a user can encounter when using an augmented reality application.}
   \label{fig:teaser}
 }

\abstract{
Recently, a lot of works show promising directions for audio design in augmented reality (AR). These works are mainly focused on how to improve user experience and make AR more realistic. But even though these improvements seem promising, these new possibilities could also be used as an input for manipulative design. This survey aims to analyze all recent discoveries in audio development regarding AR and argue what kind of ``manipulative'' effect this could have on the user. It can be concluded that even though there are many works explaining the effects of audio design in AR, very few works point out the risk of harm or manipulation toward the user. Future works could contain more awareness of this problem or maybe even systems to protect the user from any possible manipulation or harm.
} 
\keywords{Persuasive technology, Augmented reality, Virtual reality, Technology addiction, User interfaces}

\CCScatlist{
  \CCScatTwelve{Human-centered computing}{Augmented reality}{Augmented reality technologies}{};
  \CCScatTwelve{Human-centered computing}{Augmented reality}{Augmented reality design and evaluation methods}{}
}

\begin{document}



\maketitle


\section{Introduction}~\label{sec:introduction}

With the rapid development of augmented reality (AR), more detail in its design becomes important. The design of AR environments contains many influential factors like visuals, audio, haptics and wearables. When making a design for an AR environment, the setup of these influences is important to the design of the AR environment. In UX design, manipulative design occurs sometimes. In this case, the design of an interface is intentionally misleading or harming a user. So naturally, this so-called manipulative design can also occur in the design of the AR environment, as shown in Figure~\ref{fig:teaser}. But since the AR environment is a 3D environment instead of a 2D environment there are more possibilities to manipulate the design. Even though all of the aforementioned influences could affect the design of the AR environment, in this survey we study the effects of one of them: audio. This survey paper aims to research the effects of audio in manipulative design. The three concepts of audio, manipulative design and AR are the main focus of this survey.


\subsection{The impact of audio in augmented reality}
When AR, virtual reality (VR), extended reality (XR), or other immersive technologies are mentioned, people often first consider its visuals. 
Definitions of AR often describe how digital visual elements shape the technology, however, AR is not merely digital visuals.
Apart from the visuals, these technologies often make use of audio, haptics, or other sensory triggers like the perception of touch or smell. While visuals may seem the most important feature of the AR, VR, or XR virtual environment, audio might be the most developed one. Before television came, radio, the world has been using audio a lot longer in their technologies than digital visual imagery. There is a lot of information available about audio, so it can be considered that audio can be leveraged more in the case of perceptive design. 
Users are anticipated to naturally interact more with visuals 
than audio. Deceptive design of audio is harder to detect by a user than perceptive design in visuals. For this reason, it might be less obvious to the user.
Making this even a bigger threat to user safety when used improperly.

Audio in AR has a big impact on how reliable a design is, however it has another significant effect. The audio can be used to broaden the field-of-view (FOV) of the user. 
Marquardt ~\textit{et al.}~\cite{Marquardt2020comparing} explain that current AR displays still have a very limited field of view compared to human vision. So in order to localize anything outside the FOV, audio can be a very effective way to do this. There are other ways to stimulate a user to go beyond their initial FOV, however, audio is the most developed choice. 
Apart from triggering the user to broaden the FOV, the audio cues can also make a user move into a wider perspective. Or the audio can be used to specifically make the user move in a certain direction, making it a bigger liability again for perceptive AR design. So even though, audio can improve the FOV and it brings great benefit to the design of AR, it can also be a risk factor to the user's safety. 
This is why the awareness of the effects of audio design in this environment is important among designers of AR environments.

\subsection{Recent survey papers}
Recent survey papers focus mainly on other influential factors in AR or, the audio topics are regarding other immersive technologies than AR. For example, the work of Bermejo~\textit{et al.}~\cite{bermejo2021asurvey} focuses mainly on haptics and how these technologies influence the AR environment, while this survey will be focusing on audio instead of haptics. The works of Lee~\textit{et al.}~\cite{lee2021towardaugmented} focus more on Human-City Interaction and related technological approaches, by reviewing trends in
information visualization, constrained interfaces, and embodied interaction for AR headsets. So far, no previous surveys seem to specifically target the usage of audio or auditory cues in the AR environment. So in this survey, the aim is to target these kinds of technologies specifically. In another work by Fu~\textit{et al.}~\cite{fu2022systematic}, the authors describe Virtual, Augmented, and Mixed Reality Game Applications in Healthcare. This survey paper targets multiple extended reality (XR) technologies and explains their usage in healthcare environments. However, this survey will bring new insights specifically in the area of AR and audio. Different from the work on haptics used for AR, or a focus on a broad range of interactions with AR headsets, or specific applications in a technical environment, this work will focus on AR and audio as one domain. As no other surveys have aimed in this direction yet, it can make a significant difference to combine the pieces. As many works focus on either AR or audio or anything regarding these topics, this survey can link several works and bring new knowledge to the effects of audio usage in the design of AR. With these insights, safer and 
more user-friendly designs in AR can be made, considering the risks that come with certain audio setups.


\subsection{Contribution and Structure of this survey paper}
This survey paper reviews recent works on audio, audio design and AR to enable new inside on the audio domain in AR environments. The aim of this work is to highlight the connection between audio and AR and thus the potential threats or risks to users. Furthermore, this paper attempts to answer the following research questions:
 (1) What tools or technologies are used and how are they used?
 (2) What kind of effects does audio have on the user?
 (3) Is it possible for these effects to become manipulative designs?
 (4) In what cases can manipulative audio design cause a user harm?
To answer these questions, the paper structure is as follows. 
First, it is explained what audio in AR does and how manipulation of this audio in AR can affect the user. Recent surveys are still lacking some specific components that are highlighted in this work. Then, in Section 2, the systematic review methodology of this survey is explained. The search strategy, data extraction process, the survey results and the included articles are explained. In Section 3, an explanation is given of all the necessary terms that make the topic for this survey. After this, in Section 4, it is explained how these papers contribute to the survey topic. This will be followed by Section 5, in which it is explained what the research gap is, according to the collected papers in the survey, can contribute to future works. Lastly, Section 5 concludes the survey.

\section{Methodology}~\label{sec:methodology}

To conduct the literature review systematically, the PRISMA methodology was applied as described by Page~\textit{et al.}~\cite{page2021prisma}. Each paper was individually assessed and analyzed. After the analysis phase, all papers were summarized and categorized for later reference in the discussion (Section 4).

\subsection{Search Strategy}

\bparagraph{Keywords.}
The chosen keywords were in the three dimensions of audio, manipulative design and AR. The keywords used regarding audio were as follows: ``audio", ``auditory", ``ventriloquist effect" and ``McGurk effect". To take into account the influences of manipulative design, the following keywords were used: ``manipulative design", ``dark pattern", ``spatial", ``motion sickness", ``localization", ``orientation" and ``rendering". For the third dimension, only the key words ``AR" and ``augmented reality" were used (See Table~\ref{tab:keywordtab}).

\begin{table}[t]
\caption{Keywords \& search string}
\begin{tabular}{lllll}
\cline{1-2}
\begin{tabular}[c]{@{}l@{}}- Audio keywords\\ - Manipulative \\   design keywords\\ - AR keywords\end{tabular} & \begin{tabular}[c]{@{}l@{}}- (audio/ auditory/ ventriloquist effect/ \\   McGurk effect)\\ - (manipulative design/ dark pattern/ \\   spatial/ motion sickness/ localization/ \\   orientation/ rendering)\\ - (AR/ augmented reality)\end{tabular}                                                &  &  &  \\ \cline{1-2}
Search string                                                                                                   & \begin{tabular}[c]{@{}l@{}}("audio" OR "auditory" OR "ventriloquist\\  effect" OR "McGurk effect") AND \\ ("manipulative design" OR "dark pattern"\\  OR "spatial" OR "motion sickness" OR \\ "localization" OR "orientation" OR \\ "rendering") AND ("AR" OR \\ "augmented reality")\end{tabular} &  &  &  \\ \cline{1-2}

\label{tab:keywordtab}
\end{tabular}
\end{table}

\bparagraph{\textbf{Databases.}}
The ACM Digital Library (ACM DL) and IEEE Xplore Digital Library were used to find literature related to technology and computer science. A string including all the previously mentioned keywords was created to search these two databases. The search specifications included the keywords of the string in the Title, the Abstract, or Keywords of the paper. Then, the search was further defined to a period from 2014 to 2022 for more relevant results. As a result, 
84 and 63 results came up and in the IEEE Xplore Digital library and ACM Digital Library, respectively, 
resulting in a total of 147 papers.

\bparagraph{\textbf{Limitations}}
This work mainly focuses on AR and \textit{not} on other types of immersive technologies. Because of this, in the search string only the keywords "AR" and "augmented reality", and other keywords like immersive technology, XR technology, or MR technologies were left out. In future work, the effects of audio could be explored in technologies different than AR. In this survey, the focus is on audio influences only. Other surveys covered other influences in AR. Thus, this survey specifically focuses on the influence of audio in AR.

\subsection{Data Extraction}

As previously mentioned, papers would be filtered to cover the years between 2014 and 2022, this is to ensure the papers included are up to date and relevant. 
However, some of the papers may be excluded from this filter given their significance and relevance. The duplicates from the 147 results were deleted, leaving 143 papers to screen. These 143 results were screened by checking their title and abstract. If the title contained some of the keywords and the abstract included information that seemed relevant to the subject of the survey, the paper was collected and saved for later. The remaining results could be fully reviewed afterwards.

\subsection{Survey Results}
\begin{figure}
        \includegraphics[width=\columnwidth]{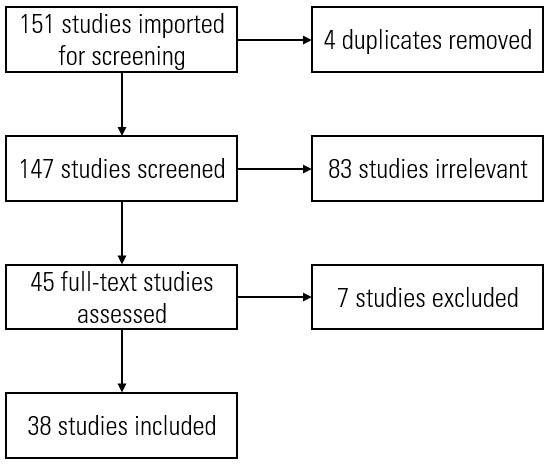}
        \caption{The process of data collection according to the PRISMA method mentioned by Page~\textit{et al.}~\cite{page2021prisma}}
        \label{fig:PRISMA}
    \end{figure}

\begin{figure}[t]
        \includegraphics[width=\columnwidth]{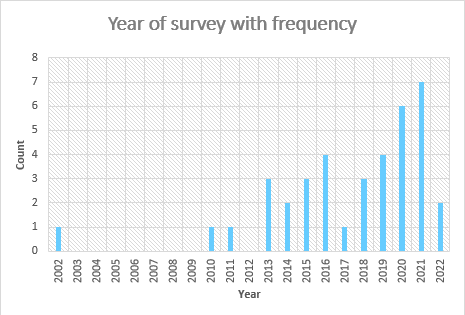}
        \caption{Publication year of articles}
        \label{fig:year}
    \end{figure}

For the outline of the systematic review process PRISMA, is used as mentioned in Figure~\ref{fig:PRISMA}.
After completing the search, the screening process started with 147 articles, of which 4 articles were removed as duplicates. The title and abstract screening process was conducted for 143 articles, of which 98 were excluded. After the full-text assessment, another 7 articles were excluded. The justifications and the counts (in parenthesis) for excluding the 7 papers are as follows: focused too much on VR rather than AR (4), focuses more on a cinematic experience rather than an AR or AR-related experience (1), does not focus on any kind of immersive technology (1) and is a poster, not a paper (1). Finally, this resulted in 38 being used for data extraction (See Figure~\ref{fig:year}).

\bparagraph{\textbf{Publication year of survey articles.}} From these remaining 38 papers, relevant information was extracted for further analysis. The information analysis was summarized 
graphically. Further qualitative insights were brought up in the iterative analysis process. The chosen set of articles dates from the years 2002 until 2022 with a peak of papers from the years 2020 and 2021. The number of publications per year can be seen in Figure~\ref{fig:year}. In the figure can be seen that there is an increase in articles regarding this survey's topic, however, this could also be caused by the growing accessibility of AR technologies. As the data extraction took place in mid-2022, the results for that year are not representative for the whole year.

\begin{figure}[t]
        \includegraphics[width=\columnwidth]{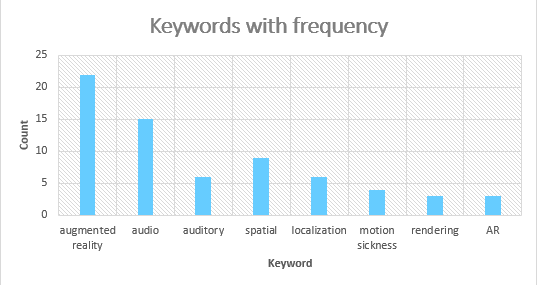}
        \caption{Frequency of keywords in articles}
        \label{fig:keyword}
    \end{figure}

\bparagraph{\textbf{Frequency of keywords in articles.}} The articles included in the survey come from many different scientific disciplines. Apart from computer science, other disciplines are medical, systems and educational sciences, etc. Most frequent keywords mention ``augmented reality", ``audio", and ``auditory" as shown in Figure~\ref{fig:keyword}. Only keywords with a frequency of 3 or higher are shown in this figure. The figure shows that the keywords relating to augmented reality and audio in some ways seem to have taken the most coverage in this set of articles. This could be the effect of the manipulative design or any keywords relating to it not being taken as a threat to the users yet. The papers focus mostly on exploring what audio can do in augmented reality rather than its effect on the user(s).

\subsection{Description of the included articles}
The included articles contain mostly papers regarding spatial audio, audio localization, audio causing motion sickness, and auditory feedback. These topics are all highly related to the effect of certain designs in AR. Below some of these works are highlighted to explain the topic.

\begin{figure}[t]
        \includegraphics[width=\columnwidth]{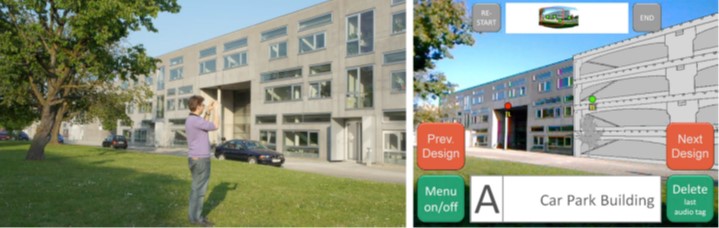}
        \caption{Example of the usage of spatial audio as mentioned in the work by Langlotz~\textit{et al.}~\cite{langlotz2013audio}}
        \label{fig:Picture1.emf}
    \end{figure}

\bparagraph{\textbf{Spatial audio.}}
Grani~\textit{et al.}~\cite{grani2015spatial} describe the effects of spatial audio in the creation of (interactive) spatial audio experiences for immersive augmented reality scenarios, and the production and mixing of the spatial audio, e.g., cinema and music. In addition to this, Langlotz~\textit{et al.}~\cite{langlotz2013audio} describe a system where audio is used spatiotemporally to find a parked car, see the example in Figure~\ref{fig:Picture1.emf}. 
The works of Hansung~\textit{et al.}~\cite{hansung2019immersive}, ~\cite{mattia2021vision} and Jot~\textit{et al.}~\cite{jot2021immersive} describe the influence of audio in an AR (/immersive) environment to estimate acoustics, room reverberation and obstacles. They explain its importance in creating a realistic AR (or other immersive) environment.

\begin{figure}[t]
        \includegraphics[width=\columnwidth]{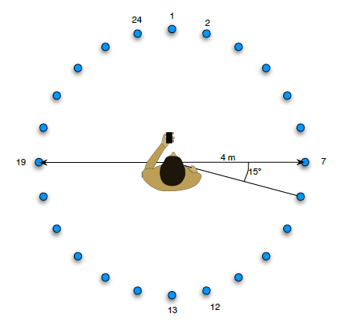}
        \caption{Example of the usage of audio localization as mentioned in the work of Heller~\textit{et al.}~\cite{heller2016where}}
        \label{fig:Picture2}
    \end{figure}

\bparagraph{\textbf{Audio localization}}
Zotkin ~\textit{et al.}~\cite{zotkin2004rendering} describe how virtual audio scene rendering is required for emerging virtual and augmented reality applications, perceptual user interfaces, and sonification of data.
Another work by Larsen~\textit{et al.}~\cite{larsen2013differences} focuses more on a common implementation of spatial audio and a head-related transfer function (HRTF) system implementation in a study in relation to precision, speed, and navigational performance in localizing audio sources in a virtual
environment rather than spatial audio.
And Binetti~\textit{et al.}~\cite{binetti2021using} explain more of the effectiveness of certain visual cues. They explain that effectiveness might be reduced if they are placed at a different visual depth plane to the target they are indicating. To overcome this visual-depth problem, we test the effectiveness of adding simultaneous spatialized auditory cues that are fixed at the target’s location. In addition to this, Heller~\textit{et al.}~\cite{heller2016where} also describe similar methods. An example of their experimental setup for audio localization is shown in Figure~\ref{fig:Picture2}.

\begin{figure}[t]
        \includegraphics[width=\columnwidth]{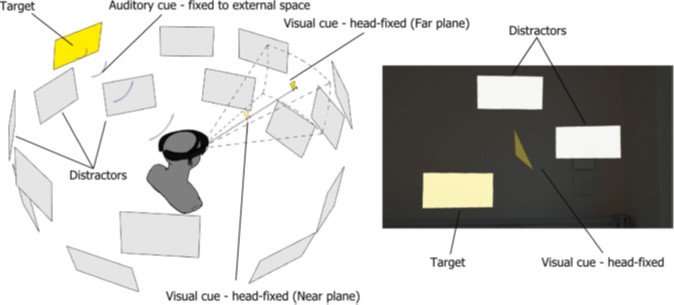}
        \caption{Example of an audio setup that was used to test if audio can create motion sickness in AR as mentioned in the work of Kuiper~\textit{et al.}~\cite{kuiper2020knowing}}
        \label{fig:Picture3.emf}
    \end{figure}

\bparagraph{\textbf{Motion sickness because of audio.}}
Dicke~\textit{et al.}~\cite{dicke2009simulator} mainly focus on motion sickness and explain that upcoming motion is known to potentially mitigate sickness resulting from provocative motion. They experimented to see what kind of effects auditory cues had that might evoke motion sickness.
Also, Kuiper~\textit{et al.}~\cite{kuiper2020knowing} explore the effect of movement patterns in a spatial sound space on the perceived amount of  simulator sickness, the pleasantness of the experience, and the perceived workload, an example of this kind of audio setup from this work is showed in Figure~\ref{fig:Picture3.emf} 

\bparagraph{\textbf{Auditory feedback.}}
Liang~\textit{et al.}~\cite{liang2020characterizing} explain that although acoustic features can capture complex
semantic information about human activities and context through human-centered mobile applications and in smart assistants in the home, continuous audio recording could pose significant privacy concerns. An intuitive way to reduce privacy concerns is to degrade audio quality such that speech and other relevant acoustic markers become unintelligible, but this often comes at the cost of activity recognition performance. 
The work of Marquardt~\textit{et al.}~\cite{Marquardt2019nonvisual} uses audio and vibrotactile feedback to guide search and information localization so that users can be better guided through with high accuracy using audio-tactile feedback
with specific settings.
And lastly, the work of Yang~\textit{et al.}~\cite{yang2019audio} explores audio augmented reality (AAR) that augments objects with 3D
sounds, which are spatialized virtually but are perceived as originating from real locations in space. The goal of their research is to enhance the people’s intuitive and immersive interactions with objects in various consumer and industrial scenarios.

\section{Taxonomy}~\label{sec:recentworks}

In this section, the taxonomy relating to the survey is explained. It is explained what audio effects are in AR, what manipulative design in AR is, and what a combination of these two could be. This is done to create an overview of the current applications of audio in AR and to expose the research gap that currently exists regarding the topic of this survey.

\subsection{Audio effects in AR}
Audio is an example of an influence in AR, and with a rapidly growing need for AR, an improvement of audio cues in AR becomes more necessary. The recent works as mentioned before in Section 2 explain the need for good audio design by explaining the impact of spatial audio and audio localization in AR. Some (bad) setups for audio in AR could even result in motion sickness as also explained in the works mentioned in Section 2. This indicates that a proper audio setup can contribute to a good user experience.

\subsection{Manipulative design in AR}
But as much as a proper audio setup can contribute to a good user experience, a bad audio setup can also contribute to manipulations in AR. Recent work~\cite{ar_advertising, gray2018dark, mathur2019dark} explains the effect of so-called dark patterns. We can find dark patterns in many scenarios during our daily lives, such as websites that request users' consent to use cookies on their site~\cite{bermejo2021website,machuletz2020multiple}. Brignull~\cite{darkPatterns} collects on his website a collection of real-world examples of dark patterns commonly used in websites and mobile apps. However, so far, very little research exists regarding this sort of manipulation in AR. In this survey, this concept is applied to the idea that audio can affect the design and can be manipulated just like in the case of dark patterns.

AR encompasses the technologies that attempt to enhance physical environments using virtual graphics, such as objects and interfaces, through the most common AR devices, including devices, wearables and haptics~\cite{chatzopoulos2017mobile}. The enriched sensing capability of AR could expose users to significant privacy threats~\cite{roesner2014security}. Thus, attackers with MAR (e.g., developers and designers) can leverage the enriched sensing capability to maximize the collection of personal data from both users and bystanders.

Mhaidli~\textit{et al.}~\cite{mhaidli2021identifying} analyze the possibilities of manipulative advertisements in XR (eXtended Reality: AR, VR, mixed reality (MR))~\cite{stanney2021extended,lee2021adcube,greenberg2014dark}. Moreover, AR technologies can also pose threats to user safety \cite{roesner2014security,jung2018ensuring,pierdicca2020augmented}. The ubiquity of MAR apps creates new opportunities for new variants of dark patterns. Our paper features an analysis of the mixed combination of virtual and physical objects in MAR scenarios, which highlight novel malicious designs, and thus new threats to individuals' privacy and safety.

\subsection{Effects of manipulative audio design in AR}
Although audio might not seem to be that important to the AR environment, a good audio setup could have a very big impact. These settings are key to guiding the user well throughout the task they are trying to accomplish in the AR environment, while also contributing to making the whole experience look more realistic. This survey paper leverages the collected research papers 
to expose those audio settings that could be an influential factor in manipulating the user and thus degrading the user experience in AR. 
This survey paper addresses existing works in the audio domain and their influences on AR, as well as the potential manipulation threat and the current research gap.

\section{Discussion}~\label{sec:discussion}

In the description of the included articles, as mentioned in Section 2.4, there is a lot of information to find regarding certain topics/ key interests for this survey paper subject. This section describes how these included articles can contribute to the research of the effects of audio manipulation in AR.

\subsection{Audio effects creating motion sickness}
One of the major things that can cause harm in AR is the audio settings that can cause motion sickness. As described in the previously mentioned articles and the other included articles
~\cite{dooley2013significant}
~\cite{Marquardt2019nonvisual}
these audio settings can cause harm to the user. The setting will make them feel stressed and disorientated and in some very bad cases, it will make the user feel nauseous or sick. This setting can be used to manipulate the user. The effect of the audio setting could be used to manipulate the design in AR. Designers could make a setup in audio to disorientate the user to trick them into doing a certain action. However, even though many works mentioned before use these kinds of effects of audio settings, so far no works seem to mention how these kind setting could be used to manipulate the user in AR. So far, we are aware of the effects of audio setups that result in motion sickness, however, it is yet to apply in an AR setting. This could have different results and maybe even other audio setups than the ones already tested could cause motion sickness in AR. This kind of setup could be dangerous and harmful to the user, even though the designers of AR are not aware of this. New research regarding this topic could make a significant contribution to developing audio design in AR.

\subsection{Localization of audio}
Although the localization of audio might not look as harmful as the motion sickness, it is still very likely to contribute significantly to the audio manipulation. As the previously mentioned studies and other included studies
~\cite{russell2016hearthere, heller2016where,
cardenas2021reducing,
miao2021investigating,
tang2020scene,
tepljakov2016sound,
chon2019matter,
kyoto2015ventri}
showed, users are able to locate audio very precisely with not too much deviation. If 
users are able to locate audio well, it means that users are also likely to be distracted by sounds coming in from different angles. Certain sounds could lure a user towards a position 
that might be different from the position a user intended to go to in the first place. Considering such scenarios, the design of audio placement could be really important to the AR environment. In the case of manipulation, we could also say that this audio setup could be used to manipulate the users' movements in the AR environment. Even though in this case the user is not physically harmed as poorly as in the case of motion sickness, this method could be much more deceitful and less noticeable to the user. Following audio can feel like a natural thing to do or even feel like an instinct, and might not seem that obvious to a user. This kind of manipulative design could very well have a big impact on the user, without the user even knowing it.

\subsection{Spatial audio}
The same idea of manipulation can be applied to the concept of motion sickness caused by spatial audio and the concept of audio localization in spatial audio. The work of Langlotz~\textit{et al.}~\cite{langlotz2013audio} describes how narrow users can trace back a car by following audio cues. 
In the case of a scenario where the user is not situated  in a small environment, the user still seems to be able to locate the audio. And in the case of motion sickness, the work of Dicke~\textit{et al.}~\cite{dicke2009simulator} explains how the user suffers from a sound circling from left to right in the user's AR environment. Both cases also show that audio in different audio spaces can still be affected by poor audio design. Other works

~\cite{baldwin2017scatar,
kailas2021design,
cokelek2021leveraging,
erkut2018mobile,
ruminski2015modeling,
vazquez2012auditory}
also describe similar situations that explain the effects of audio settings in spatial environments. 
These works talk about how easy it is to change audio, but they don't talk about what kind of effects these different audio settings could have in AR yet.

\subsection{Other included articles}
Others works mention the application of audio influences in the guidance of museums ~\cite{heller2016where}. Another work ~\cite{marto2020multisensory} mentions the effects of different stimuli in AR environments. In addition to audio and visuals, they also mention the effect of smell and results show that combinations of visuals and other stimuli are a lot more attractive to the user than just the visuals. This is also a very interesting application to the real-world usage of AR.
Another work mentions the manipulation of voice tones~\cite{wang2018effect}. This could be a very important factor in the usage of virtual assistants. When the virtual assistant guides the user, the voice tone and pitch might have a big influence on the users' interactions. 
And lastly, the work of ~\textit{et al.}~\cite{Moustakas2020adaptive} mentions the effect of audio mixing in audio AR games. In their results, it is shown that the mixes that were provided gave better game performance compared to the cases that did not include the audio mix. These papers all describe effects that could be quite important when it comes to future work in manipulative audio design in AR.

\subsection{Preventive framework}

As previously mentioned in Section 1, audio can have different kinds of effects on the usage of AR. It can be used to enhance the FOV, but it can also be used to attract the user to the usage of AR. It can be used to enhance the FOV, but it can also be used to attract the user to move in a specific direction. In the previous sections, it was explained how audio can be localized by users, how audio can be used spatially, and how audio can be used to create motion sickness among users. Although it has not yet been mentioned in the works, audio can be used to deceive users of an AR environment. To prevent this kind of situation, and protect the users, the design of audio in an AR environment has to be thoroughly considered. To be able to do so, the following framework can be presented. Designers have to consider the following effect of audio in their design:

 \begin{itemize}
     \item \textbf{Audio effects that cause motion sickness.} Some specific setups of audio might be very harmful to the user. A new potential design should be compared to other designs that resulted in motion sickness, to determine if the new design will result in motion sickness.
     \item \textbf{Audio effects that can be used to localize audio.} Audio can be used as a means to localize an object or to make a user change its FOV. This setup for the audio in the AR can be used as an effective way to redirect the user's behavior, but it can also be used to trick a user into doing something that they might not have intended to do. To intercept this design flaw, designers should consider their setup thoroughly. Audio should not be used so that a user can localize sound, to lure them into moving towards a direction they never planned on going to. If the designer considers this concept and makes a design according to this guideline, there will be no conflict or danger to the user.
     \item \textbf{Audio effects that can be used spatially.} Users can follow the audio either by following some audio cues that are given. This also creates an environment that is sensitive to manipulative design. Too many audio cues might overwhelm a user, and traveling audio might be hard to follow for a user. The number of audio inputs should therefore be limited in a design, and the number of iterations in audio design in an AR environment should also be limited. This is to protect the user and make sure that the audio is not misused to make the user move to or from specific locations in the AR environment. 
 \end{itemize}
 
It can be suggested to create some ethical design guidelines based on the abovementioned points. As such, users can make use of AR in a safe and harm-free environment.

\section{Research roadmap}~\label{sec:research_roadmap}
A research roadmap and the corresponding open questions to support this survey topic can be proposed after analyzing the existing literature in Sections 2 and 3.

\subsection{Current research gap}
\begin{figure}
        \includegraphics[width=0.85\columnwidth]{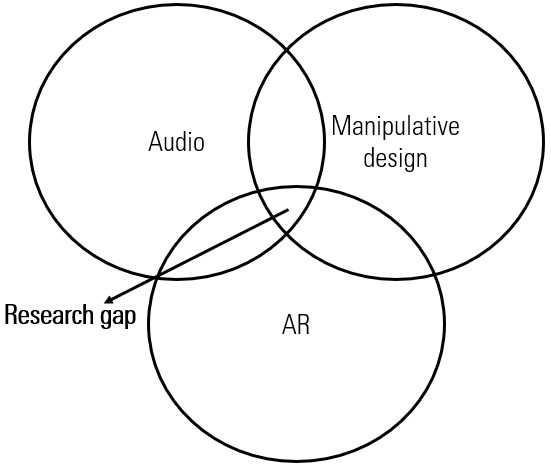}
        \caption{The research gap between audio, manipulative design and AR visualized}
        \label{fig:RESEARCHGAP}
    \end{figure}

This field of research is aware of the importance of audio technologies like AR, VR, and XR. So far the research shows that we can influence the user's behavior and make a user change its FOV, but not how the audio setup in AR specifically affects a user. As mentioned in Section 3, recent works primarily describe the importance of good audio setup in AR; however, no work has yet clearly explained what would happen if these methods were used to manipulate audio setup. This results in a research gap, as shown in Figure~\ref{fig:RESEARCHGAP}. Most of the effects of the design in neutral settings are explained in the works on how sound effects can cause motion sickness, how sound can be localized, and how sounds
spread in a user environment.
Specifically, works on audio localization and spatial audio rarely explain the "negative" effects of audio settings or how they can harm the user. The works on motion sickness explain the risk to the user, but they make no mention of how such audio setups could manipulate the user in any way.

\subsection{Opportunities for future works}
 There is a gap from research about the effects of (manipulative) audio setup to the effects of (manipulative) AR design. This could be done to develop new design guidelines and to protect the user from any harm or undesired influences when moving through the AR environment. It is important to understand what could be manipulated in audio design to make any kind of guideline to protect the user. This survey summarizes all of these works and current technologies, allowing one to conduct research and attempt to find a way to protect users from harm or manipulative designs. 

\section{Conclusion}~\label{sec:methodology}
This article surveys the area of (manipulative) audio effects in AR. Specifically, this survey summarized and categorized the current tools and technologies (mostly regarding motion sickness caused by audio, audio localization, and spatial audio). This paper gives a summary of the known effects on the user and talks about how to bridge the gap to find out what else these effects do. The survey also explored the possibilities of these audio effects being turned into manipulative designs in the AR environment. The survey article ends by presenting a research roadmap and several open research questions regarding the possible effects of manipulative audio design in an AR environment and why it is important to develop a protection system for this to keep the user safe. 



\bibliographystyle{abbrv-doi}

\bibliography{review_audio_ar.bib}

\end{document}